# Charge, neutron, and weak size of the atomic nucleus

G. Hagen[1,2], A. Ekström[1,2], C. Forssén[1,2,3], G. R. Jansen[1,2], W. Nazarewicz[1,4,5], T. Papenbrock[1,2], K. A. Wendt[1,2], S. Bacca[6,7], N. Barnea[8], B. Carlsson[3], C. Drischler[9,10], K. Hebeler[9,10], M. Hjorth-Jensen[4,11], M. Miorelli[6,12], G. Orlandini[13,14], A. Schwenk[9,10] & J. Simonis[9,10]

**Summary**

What is the size of the atomic nucleus? This deceivably simple question is difficult to answer. While the electric charge distributions in atomic nuclei were measured accurately already half a century ago, our knowledge of the distribution of neutrons is still deficient. In addition to constraining the size of atomic nuclei, the neutron distribution also impacts the number of nuclei that can exist and the size of neutron stars. We present an *ab initio* calculation of the neutron distribution of the neutron-rich nucleus $^{48}$Ca. We show that the neutron skin (difference between radii of neutron and proton distributions) is significantly smaller than previously thought. We also make predictions for the electric dipole polarizability and the weak form factor; both quantities are currently targeted by precision measurements. Based on *ab initio* results for $^{48}$Ca, we provide a constraint on the size of a neutron star.

[1]Physics Division, Oak Ridge National Laboratory, Oak Ridge, Tennessee 37831, USA. [2]Department of Physics and Astronomy, University of Tennessee, Knoxville, Tennessee 37996, USA. [3]Department of Fundamental Physics, Chalmers University of Technology, SE-412 96 Göteborg, Sweden. [4]Department of Physics and Astronomy and NSCL/FRIB, Michigan State University, East Lansing, Michigan 48824, USA. [5]Faculty of Physics, University of Warsaw, Pasteura 5, 02-093 Warsaw, Poland. [6]TRIUMF, 4004 Wesbrook Mall, Vancouver, British Columbia, Canada V6T 2A3. [7]Department of Physics and Astronomy, University of Manitoba, Winnipeg, Manitoba, Canada R3T 2N2. [8]Racah Institute of Physics, Hebrew University, 91904 Jerusalem, Israel. [9]Institut für Kernphysik, Technische Universität Darmstadt, 64289 Darmstadt, Germany. [10]ExtreMe Matter Institute EMMI, GSI Helmholtzzentrum für Schwerionenforschung GmbH, 64291 Darmstadt, Germany. [11]Department of Physics, University of Oslo, N-0316 Oslo, Norway. [12]Department of Physics and Astronomy, University of British Columbia, Vancouver, British Columbia, Canada V6T 1Z4. [13]Dipartimento di Fisica, Universita di Trento, I-38123 Trento, Italy. [14]Istituto Nazionale di Fisica Nucleare, TIFPA, I-38123 Trento, Italy.

Atomic nuclei are made of two types of fermions – protons and neutrons. Due to their electric charge, the distribution of protons in a nucleus can be accurately measured and is well known for many atomic nuclei[1]. In contrast, neutron densities are poorly known. An accurate knowledge of neutron distributions in atomic nuclei is crucial for understanding neutron-rich systems ranging from short-lived isotopes at the femtometer scale to macroscopically large objects such as neutron stars. The distribution of neutrons in nuclei determines the limits of the nuclear landscape[2], gives rise to exotic structures and novel phenomena in rare isotopes[3-5], and impacts basic properties of neutron stars[6-8]. Because of its fundamental importance, experimental efforts worldwide have embarked on an ambitious program of measurements of neutron distributions in nuclei using different probes, including hadronic scattering[9], pion photoproduction[10], and parity-violating electron scattering[11]. Electrons interact with nucleons by exchanging photons and $Z^0$ bosons. Since neutrons have no electric charge, elastic electron scattering primarily probes the proton distribution. On the other hand, parity-violating electron scattering can only occur via the weak interaction and is sensitive to the distribution of weak charge. As the weak charge of the neutron, $Q_W^n \approx -0.99$, is much larger than that of the proton, $Q_W^p \approx 0.07$, a measurement of the parity violating asymmetry $A_{pv}$ (ref. 12) offers an opportunity to probe the neutron distribution.

Regardless of the probe used, direct measurements of neutron distributions in nuclei are extremely difficult. For this reason, experiments have also focused on other observables related to neutron distributions, such as the electric dipole polarizability $\alpha_D$. Recently, $\alpha_D$ was accurately determined in $^{208}$Pb (ref. 13), $^{120}$Sn (ref. 14) and $^{68}$Ni (ref. 15), while an experimental extraction of $\alpha_D$ for $^{48}$Ca by the Darmstadt-Osaka collaboration is ongoing. For this medium-mass nucleus, the Calcium Radius Experiment (CREX) at Jefferson Lab[16] also aims at a measurement of the radius of the weak charge distribution. The nucleus $^{48}$Ca is of particular interest because it is neutron rich, has doubly-magic structure, and can now be reached by nuclear *ab initio* methods.

So far, much of the theoretical understanding of proton and neutron distributions in atomic nuclei came from nuclear density functional theory (DFT)[17]. This method employs energy density functionals that are primarily constrained by global nuclear properties such as binding energies and radii, and it provides us with a coarse-grained description of nuclei across the nuclear chart. Calculations within nuclear DFT generally describe charge radii, and suggest that $\alpha_D$ is strongly correlated with the neutron skin[18-20], thereby relating this quantity to the neutron radius. To be able to tackle a medium-mass nucleus such as $^{48}$Ca with both *ab initio* and DFT methods provides an exciting opportunity to bridge both approaches. In the process, surprises are expected. For instance, as discussed in this work, *ab initio* calculations show that the neutron skin of $^{48}$Ca is significantly smaller than estimated by nuclear DFT models. This result not only gives us an important insight into the nuclear size, but also provides an opportunity to inform global DFT models by more refined *ab initio* theories.

In recent years, *ab initio* computations of atomic nuclei have advanced tremendously. This progress is due to an improved understanding of the strong interaction that binds protons and neutrons into finite nuclei, significant methodological and algorithmic advances, and ever-increasing computer performance. In this work, we use nuclear forces derived from chiral effective field theory[21,22] that are rooted in quantum chromodynamics, the theory of the strong interaction. The quest for nuclear forces of high fidelity has now reached a critical stage (Fig. 1a).

In this study we use the recently developed next-to-next-to-leading order chiral interaction NNLO$_{sat}$ (ref. 23) that is constrained by radii and binding energies of selected nuclei up to mass number $A\approx25$. It provides a basis for accurate *ab initio* modeling of light and medium-heavy nuclei. Combined with a significant progress in algorithmic and computational developments in recent years[24], the numerical cost of solving the *ab initio* nuclear many-body problem has changed from exponential to polynomial in the number of nucleons $A$, with coupled-cluster theory being one of the main drivers[24]. The present work pushes the frontier of accurate nuclear *ab initio* theory all the way to $^{48}$Ca (Fig. 1b).

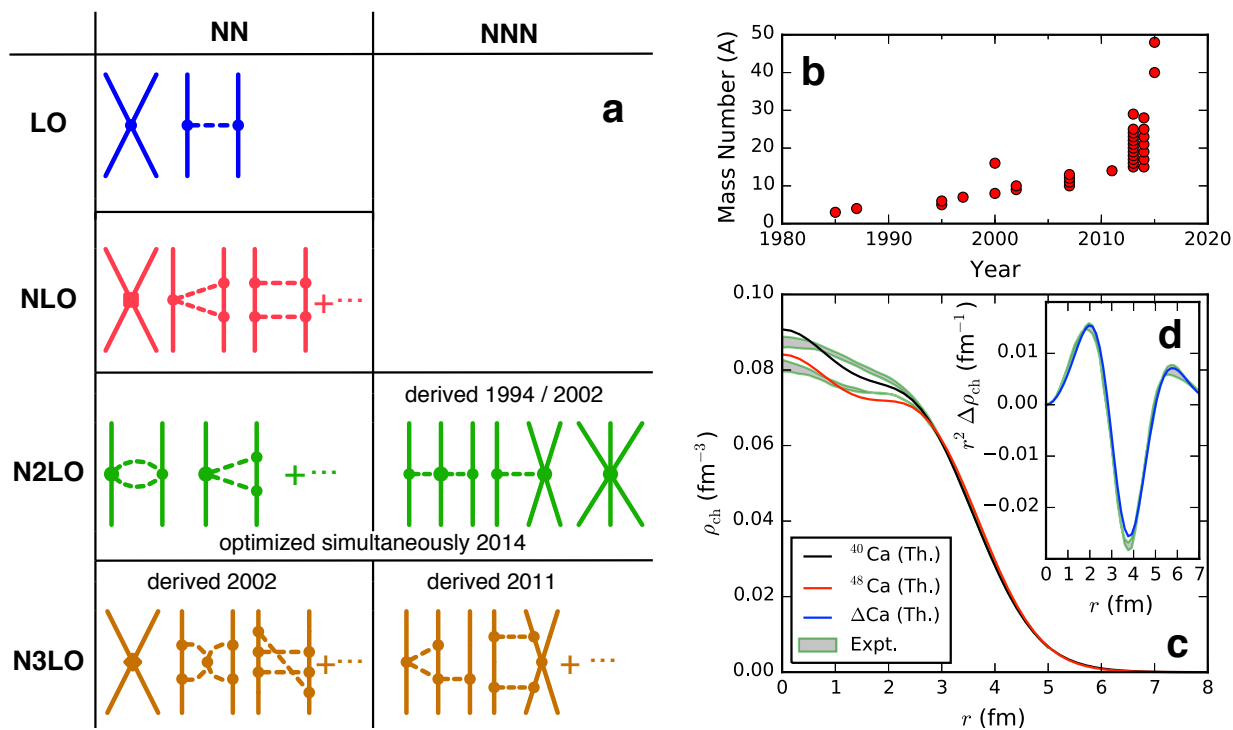

**Figure 1 | A*b initio* computations for atomic nuclei. a**, Diagrammatic illustration of nuclear forces based on chiral effective field theory[21,22], with nucleons being shown as full lines and exchanged pions as dashed lines. The left column corresponds to nucleon-nucleon (NN) interactions, while the right column shows three-nucleon (NNN) diagrams. Rows display contributions from diagrams of leading order (LO), next-to-leading order (NLO), etc.; progress milestones are indicated. **b**, Trend of realistic *ab initio* calculations for the nuclear *A*–body problem. In the early decades, the progress was approximately linear in the mass number *A* because the computing power, which increased exponentially according to the Moore's law, was applied to exponentially expensive numerical algorithms. In recent years, however, new-generation algorithms, which exhibit polynomial scaling in *A,* have dramatically increased the reach. **c**, *Ab initio* predictions (this work) for charge densities in $^{40}$Ca (black line) and $^{48}$Ca (red line) compared to experiment[26] (shaded area). **d**, The difference between the computed charge densities of $^{40}$Ca and $^{48}$Ca (blue line) compared to experiment (shaded area).

Our NNLO$_{sat}$ predictions for the electric charge densities $\rho_{ch}$ in $^{40}$Ca and $^{48}$Ca are shown in Figure 1c, see the Methods section for details. The agreement of theoretical charge densities with experiment[25], especially in the surface region, is most encouraging. The difference between charge densities of $^{40}$Ca and $^{48}$Ca (shown in the inset of Fig. 1c) is even better reproduced by theory as systematic errors at short distances cancel out. The striking similarity of the measured charge radii of $^{40}$Ca and $^{48}$Ca, 3.478(2) fm and 3.477(2) fm, respectively, has been a long-standing challenge for microscopic nuclear structure models. Our results for the charge radii are 3.49(3) fm for $^{40}$Ca and 3.48(3) fm for $^{48}$Ca; these are the first *ab initio* calculations to successfully reproduce this observable in both nuclei. The distribution of the electric charge in a nucleus profoundly impacts the electric dipole polarizability. To compute this quantity, we have extended the formalism of ref. 26 to accommodate three-nucleon forces. In order to validate our model, we computed the dipole polarizabilities of $^{16}$O and $^{40}$Ca, for which experimental data exist[27]. We find an excellent agreement with experiment for $^{16}$O, $\alpha_D = 0.57(1)$ fm$^3$ compared to $\alpha_{D,exp} = 0.58(1)$ fm$^3$. Our result for $^{40}$Ca, $\alpha_D = 2.11(4)$ fm$^3$, is only slightly below the experimental value $\alpha_{D,exp} = 2.23(3)$ fm$^3$.

We now turn to our main objective and present our predictions for the point-neutron radius (i.e., the radius of the neutron distribution) $R_n$, point-proton radius $R_p$, neutron skin $R_{skin} = R_n - R_p$, and electric dipole polarizability in $^{48}$Ca. Point radii are related to the experimentally measured (weak-) charge radii by corrections that account for the finite size of the nucleon (see the Methods section for details). To estimate systematic uncertainties on computed observables, in addition to NNLO$_{sat}$, we consider a family of chiral interactions[28]. Similar to NNLO$_{sat}$, these interactions consist of soft nucleon-nucleon and non-local three-nucleon forces. Their three-nucleon forces were adjusted to the binding energy of $^3$H and the charge radius of $^4$He only, and – within EFT uncertainties – they yield a realistic saturation point of nuclear matter[28], and reproduce two-neutron separation energies of calcium isotopes[4], see Extended Data Table 2. A main difference between these interactions and NNLO$_{sat}$ is that they have not been constrained by experimental data on heavier nuclei, and they include next-to-next-to-next-to leading order nucleon-nucleon contributions.

Figure 2 shows the predicted neutron skin, point-neutron radius, and dipole polarizability as functions of the point-proton radius. In all three panels of Fig. 2, the blue line represents a linear fit to our *ab initio* results obtained with the set of chiral forces considered. The blue bands provide an estimate of systematic uncertainties (see Methods section). They encompass the error bars on the computed data points and are symmetric around the linear fit (blue line). The charge radius of $^{48}$Ca is known precisely, and the horizontal green line marks the corresponding point-proton radius $R_p$. The intersection between this line and the blue band provides a range for these observables (shown as vertical orange bands) consistent with our set of interactions. Our prediction for the neutron skin in $^{48}$Ca is $0.12 \lesssim R_{skin} \lesssim 0.15$ fm. Figure 2a shows two remarkable features. First, the *ab initio* calculations yield neutron skins that are almost independent of the employed interaction. This is due to the strong correlation between the point-neutron and point-proton radii in this nucleus (Fig. 2b). In contrast, DFT models exhibit practically no correlation between the neutron skin and the point-proton radius. Second, the *ab initio* calculations predict a significantly smaller neutron skin than the DFT models. The predicted range is also appreciably lower than the combined DFT estimate of 0.176(18) fm (ref. 19) and is well below the relativistic DFT value of $R_{skin}$=0.22(2) fm (ref. 19). To shed light on the lower values of $R_{skin}$ predicted by *ab initio* theory, we computed the neutron separation energy and the three-point binding energy difference in $^{48}$Ca (both being indicators of the $N$=28 shell gap). Our results are consistent with experiment and indicate the pronounced magicity of $^{48}$Ca (Extended Data Table 2), while DFT results usually significantly underestimate the $N$=28 shell gap[29]. The shortcoming of DFT for $^{48}$Ca is also reflected in the point-proton radius. Although many nuclear energy density functionals are constrained to the point-proton radius of $^{48}$Ca[17,29], the results of DFT models shown in Fig. 2a overestimate this quantity.

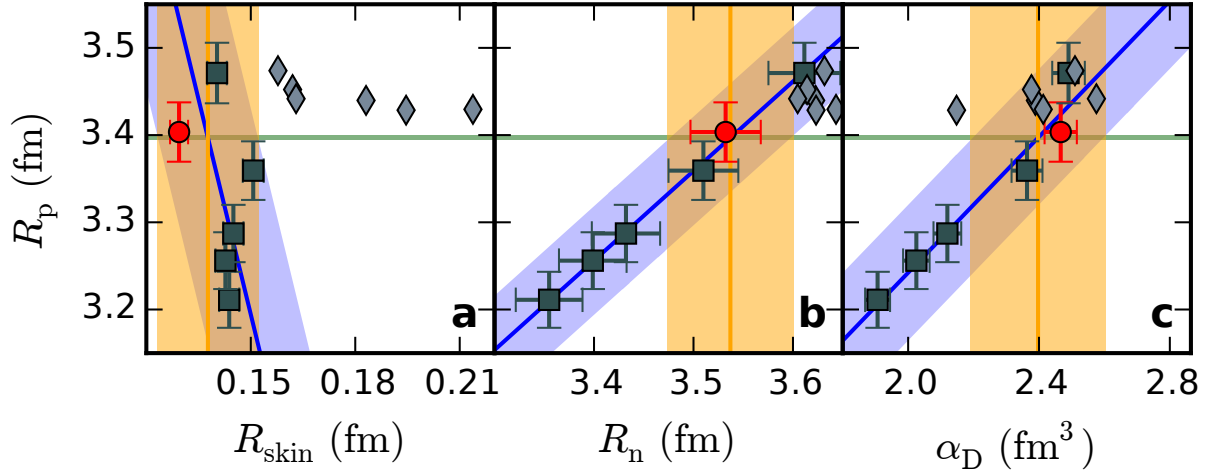

**Figure 1 | Predictions for observables related to the neutron distribution in $^{48}$Ca. a**, the neutron skin $R_{skin}$; **b**, the point-neutron radius $R_n$; and **c**, the electric dipole polarizability $\alpha_D$ – all versus the point-proton radius $R_p$. The *ab initio* predictions with NNLO$_{sat}$ (dots) and chiral interactions of ref. 28 (squares) are compared to the DFT results with the energy density functionals SkM*, SkP, SLy4, SV-min, UNEDF0, and UNEDF1[19] (diamonds). The theoretical error bars are indicated. The blue line represents a linear fit to the data, with theoretical uncertainties shown by a blue band. The horizontal green line marks the experimental value of $R_p$ that puts a constraint on the ordinate (orange band).

For the point-neutron radius (Fig. 2b) we find $3.47 \leq R_n \leq 3.60$ fm. Most of the DFT results for $R_n$ fall within this band. Comparing Figs. 2a and 2b suggests that a measurement of a small neutron skin in $^{48}$Ca would provide a critical test for *ab initio* models. For the electric dipole polarizability (Fig. 2c) our prediction $2.19 \leq \alpha_D \leq 2.60$ fm$^3$ is consistent with the DFT value of 2.306(89) fm$^3$ (ref. 19). Again, most of the DFT results fall within the *ab initio* uncertainty band. The result for $\alpha_D$ will be tested by anticipated experimental data from the Darmstadt-Osaka collaboration[13,14]. The excellent correlation between $R_p$, $R_n$, and $\alpha_D$ seen in Figs. 2b and 2c demonstrates the usefulness of $R_n$ and $\alpha_D$ as probes of neutron density.

The weak charge radius $R_W$ is another quantity that characterizes the size of the nucleus. The CREX experiment will measure the parity violating asymmetry $A_{pv}$ in electron scattering on $^{48}$Ca at the momentum transfer $q_c=0.778$ fm$^{-1}$. This observable is proportional to the ratio of the weak and electromagnetic charge form factors $F_W(q_c)/F_{ch}(q_c)$ (ref. 12). Making some assumptions about the weak-charge form factor, one can deduce the weak-charge radius $R_W$ and the point-neutron radius $R_n$ from the single CREX data point[16]. Figure 3a shows that a strong correlation exists between $R_n$ and $F_W(q_c)$, and this allows us to estimate $0.195 \leq F_W(q_c) \leq 0.222$ (Extended Data Fig. 2), which is consistent with the DFT expectation[20]. The momentum dependence of the weak-charge form factor (Fig. 3b) is also close to the DFT result. This good agreement again emphasizes the role of $^{48}$Ca as a key isotope for bridging nuclear *ab initio* and DFT approaches. Exploiting the strong correlation between $R_W$ and $R_p$, we find $3.59 \leq R_W \leq 3.71$ fm (Extended Data Fig. 1). The weak-charge density $\rho_W(r)$ is the Fourier transform of the weak-charge form factor $F_W(q)$. As seen in Fig. 3c, the spatial extent of $\rho_W$ in $^{48}$Ca is appreciably greater than that of the electric charge density $\rho_{ch}$, essentially because the former depends mainly on the neutron distribution and there is an excess of eight neutrons over protons in $^{48}$Ca.

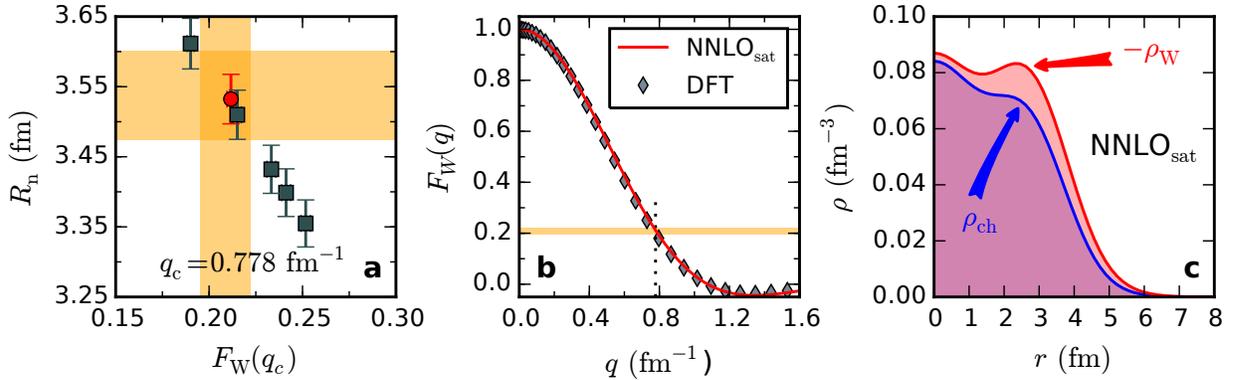

**Figure 3 | Weak-charge observables in $^{48}$Ca. a**, Point-neutron radius $R_n$ in $^{48}$Ca versus the weak charge form factor $F_W(q_c)$ at the CREX momentum $q_c= 0.778$ fm$^{-1}$ obtained in *ab initio* calculations with NNLO$_{sat}$ (red circle) and chiral interactions of ref. 28 (squares). The orange bands show the predicted ranges for $R_n$ and $F_W(q_c)$. **b**, Weak charge form factor $F_W(q)$ as a function of momentum transfer $q$ with NNLO$_{sat}$ (red line) and DFT with the energy density functional SV-min[20] (diamonds). The orange horizontal band shows $F_W(q_c)$. **c**, Charge density (blue line) and (negative of) weak charge density (red line). The weak charge density extends well beyond $\rho_{ch}$ as it is strongly weighted by the neutron distribution. The weak charge of $^{48}$Ca, obtained by integrating the weak charge density is $Q_W=-26.22$.

The neutron distribution in atomic nuclei is related to the nuclear matter equation of state, which in turn impacts the size of neutron stars[6-8]. Since the set of interactions employed in this work has turned out to be useful for gauging systematic trends of observables that depend on neutron density (see Fig. 2), this offers an opportunity to estimate the symmetry energy $S_v$ and its slope $L$ at nuclear saturation density (see Methods section). As seen in Figs. 4a and 4b, our calculations of asymmetric nuclear matter yield results for $S_v$ and $L$ that are well correlated with the point-proton radius of $^{48}$Ca. This allows us to deduce $25.2 \leq S_v \leq 30.4$ MeV, $37.8 \leq L \leq 47.7$ MeV. These estimates are consistent with the recently suggested ranges $29.0 \leq S_v \leq 32.7$ MeV and $40.5 \leq L \leq 61.9$ MeV (ref. 30). The chiral forces used in our analysis have been constrained around nuclear saturation density, which is much smaller than the actual density in the interior of a neutron star. For that reason, their straightforward extrapolations to supra-saturation densities are not supposed to be meaningful. However, there exists an empirical power law that relates neutron-star radii to the pressure $P$ at nuclear saturation density[31]. Furthermore, $P$ is strongly connected to $S_v$ and $L$ and can also be expected to correlate with the point-proton radius of $^{48}$Ca. Exploiting this correlation we arrive at an estimate $2.3 \lesssim P \lesssim 2.6$ MeV fm$^{-3}$ (see Methods section and Extended Data Fig. 3). Figure 4c shows the predicted radius $11.1 \lesssim R_{1.4M_\odot} \lesssim 12.7$ km of a $1.4M_\odot$ neutron star based on this pressure and the phenomenological expression of refs. 30,31. It is compatible with radius estimates based on high-density extensions of *ab initio* results for the equation of state[8], the analysis of ref. 30, and results from a Bayesian analysis of quiescent low-mass X-ray binaries[32]. This is a very encouraging result. In order to improve our *ab initio* description one needs to develop a well-calibrated, higher-order chiral interaction, which will extend energy, momentum, and density range of our *ab initio* framework. This is a long-term goal.

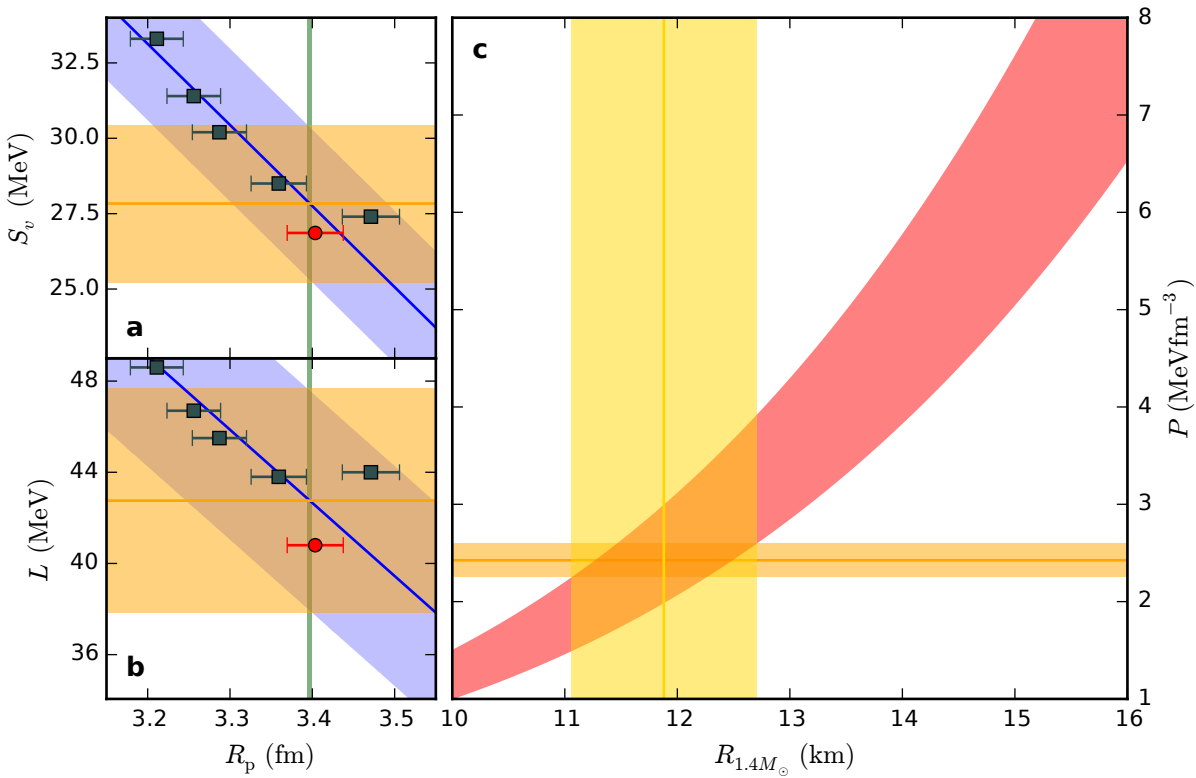

**Figure 4 | Properties of the nuclear equation of state and neutron star radii based on chiral interactions. a**, The symmetry energy $S_v$ and **b**, the slope $L$ of the symmetry energy at predicted saturation densities versus the point-proton radius in $^{48}$Ca. **c**, Pressure-radius relationship for a neutron star of mass $M=1.4M_\odot$ (red band) from the phenomenological expression of refs. 30,31. The predicted pressure (horizontal orange band) constrains the neutron star radius (vertical yellow band).

**References**

1. I. Angeli, K. P. Marinova, Table of experimental nuclear ground state charge radii: An update, *At. Data Nucl. Data Tabl.* **99**, 69 (2013).

2. J. Erler *et al.,* The limits of the nuclear landscape*, Nature* **486**, 509-512 (2012).

3. I. Tanihata *et al.,* Measurements of interaction cross sections and nuclear radii in the light p-shell region, *Phys. Rev. Lett.* **55**, 2676-2679 (1985).

4. F. Wienholtz *et al.*, Masses of exotic calcium isotopes pin down nuclear forces, *Nature* **498**, 346-349 (2013).

5. D. Steppenbeck *et al.,* Evidence for a new nuclear 'magic number' from the level structure of $^{54}$Ca, *Nature* **502**, 207-210 (2013).

6. B. A. Brown, Neutron radii in nuclei and the neutron equation of state, *Phys. Rev. Lett.* **85**, 5296-5299 (2000).

7. J. M. Lattimer, M. Prakash, The physics of neutron stars, *Science* **304**, 536-542 (2004).

8. K. Hebeler, J. M. Lattimer, C. J. Pethick, A. Schwenk, Equation of state and neutron star properties constrained by nuclear physics and observation, *Astrophys. J.* **773**, 11 (2013).

9. J. Zenihiro *et al.,* Neutron density distributions of $^{204,206,208}$Pb deduced via proton elastic scattering at Ep=295 MeV*, Phys. Rev. C* **82**, 044611 (2010).

10. C. M. Tarbert *et al.,* Neutron skin of $^{208}$Pb from coherent pion photoproduction, *Phys. Rev. Lett.* **112**, 242502 (2012).

11. S. Abrahamyan *et al.,* Measurement of the neutron radius of $^{208}$Pb through parity violation in electron scattering, *Phys. Rev. Lett.* **108**, 112502 (2012).

12. T. W. Donnelly, J. Dubach, I. Sick, Isospin dependences in parity-violating electron scattering, *Nucl. Phys. A* **503**, 589-631 (1989).

13. A. Tamii *et al*., Complete electric dipole response and the neutron skin in $^{208}$Pb, *Phys. Rev. Lett.* **107**, 062502 (2011).

14. T. Hashimoto *et al*, Dipole polarizability of $^{120}$Sn and nuclear energy density functionals, *arXiv*:1503.08321 (2015).

15. D. M. Rossi *et al*., Measurement of the dipole polarizability of the unstable neutron-rich nucleus $^{68}$Ni, *Phys. Rev. Lett.* **111**, 242503 (2013).

16. S. Riordan *et al.,* CREX proposal to Jefferson Lab (2013); C. J. Horowitz, K. S. Kumar, R. Michaels, Electroweak measurements of neutron densities in CREX and PREX at JLab, USA, *Eur. Phys. J. A* **50**: 48 (2014).

17. M. Bender, P.-H. Heenen, P.-G. Reinhard, Self-consistent mean-field models for nuclear structure, *Rev. Mod. Phys.* **75**, 121-180 (2003).

18. P.-G. Reinhard, W. Nazarewicz, Information content of a new observable: The case of the nuclear neutron skin, *Phys. Rev. C* **81**, 051303(R) (2010).

19. J. Piekarewicz *et al.*, Electric dipole polarizability and the neutron skin, *Phys. Rev. C* **85**, 041302 (2012).

20. P.-G. Reinhard *et al.,* Information content of the weak-charge form factor*, Phys. Rev. C* **88**, 034325 (2013).

21. E. Epelbaum, H.-W. Hammer, Ulf-G. Meißner, Modern theory of nuclear forces, *Rev. Mod. Phys*. **81**, 1773 (2009).

22. R. Machleidt, D. Entem, Chiral effective field theory and nuclear forces, *Phys. Rep.* **503**, 1-75 (2011).

23. A. Ekström *et al.,* Accurate nuclear radii and binding energies from a chiral interaction, *Phys. Rev. C* **91**, 051301(R) (2015).

24. G. Hagen, T. Papenbrock, M. Hjorth-Jensen, D. J. Dean, Coupled-cluster computations of atomic nuclei, *Rep. Prog. Phys.* **77**, 096302 (2014).

25. H. J. Emrich *et al.,* Radial distribution of nucleons in isotopes $^{48}$Ca, $^{40}$Ca, *Nucl. Phys. A* **396**, 401c-408c (1983).

26. S. Bacca *et al.,* Giant and pigmy dipole resonances in $^{4}$He, $^{16,22}$O, and $^{40}$Ca from chiral nucleon-nucleon interactions*, Phys. Rev. C* **90**, 064619 (2014).

27. J. Ahrens *et al.,* Total nuclear photon absorption cross-sections for some light elements, *Nucl. Phys. A* **251**, 479-492 (1975).

28. K. Hebeler, S. K. Bogner, R. J. Furnstahl, A. Nogga, A. Schwenk, Improved nuclear matter calculations from chiral low-momentum interactions, *Phys. Rev. C* **83**, 031301 (2011).

29. M. Kortelainen *et al.,* Nuclear energy density optimization: Shell structure, *Phys. Rev. C* **89**, 054314 (2014).

30. J. M. Lattimer, Y. Lim, Constraining the symmetry parameters of the nuclear interaction, *Astrophys. J.* **771**, 51 (2013).

31. J. M. Lattimer, M. Prakash, Neutron star structure and the equation of state, *Astrophys. J.* **550**, 426 (2001).

32. J. M. Lattimer, A. W. Steiner, Neutron star masses and radii from quiescent low-mass X-ray binaries, *Astrophys. J.* **784**, 123 (2014).

## Methods

**Hamiltonian and model space.**
The *ab initio* coupled-cluster calculations employ the intrinsic Hamiltonian $H = T - T_{cm} + V_{NN} + V_{3NF}$, where $T$ is the total kinetic energy, $T_{cm}$ the kinetic energy of the center-of-mass, $V_{NN}$ is the nucleon-nucleon interaction and $V_{3NF}$ is the three-nucleon force (3NF). We employ several interactions to estimate theoretical uncertainties. The interaction NNLO$_{sat}$ from chiral effective field theory (EFT) at next-to-next to leading order (NNLO) was adjusted to reproduce binding energies and radii in selected nuclei up to mass number *A*≈25 (*23*). Another set of interactions was taken from ref. 28. These interactions employ similarity renormalization group transformations[33] of the nucleon-nucleon interaction[34] at next-to-next-to-next to leading order (N3LO) from chiral EFT. The corresponding 3NF takes into account contributions at NNLO with low-energy coefficients $c_D$ and $c_E$ adjusted to binding energy of the triton and the radius of the alpha particle, see Extended Data Table 1 and ref. 28 for more details. These interactions reproduce two-neutron separation energies and spectroscopy of neutron-rich calcium isotopes[4,35]. Our single-particle basis consists of 15 major harmonic oscillator shells with an oscillator frequency of $\hbar\omega = 22$ MeV, and the 3NF is truncated to the three-particle energies with $E_{3max} \leq 18\hbar\omega$ for NNLO$_{sat}$ and $E_{3max} \leq 16\hbar\omega$ for the other chiral Hamiltonians. A Hartree-Fock calculation yields the reference state for the coupled-cluster computation. The Hamiltonian is normal-ordered with respect to the Hartree-Fock reference state, and we use the normal-ordered two-body approximation for the 3NF. As demonstrated in refs. 36,37 this approximation is precise for light and medium mass nuclei.

**Coupled-cluster method.**
The quantum nuclear many-body problem is solved with the coupled-cluster method, see ref. 24 for a recent review of nuclear coupled-cluster computations. Coupled-cluster theory performs the similarity transform $\bar{H} = e^{-T}He^{T}$ of the Hamiltonian *H* using the cluster operator *T* that consists of a linear expansion in particle-hole excitation operators. Approximations are introduced by truncating the operator *T* to a lower particle-hole rank, and the most commonly used approximation is coupled-cluster with singles and doubles excitations (CCSD). For the computation of binding energy of $^{48}$Ca we include the perturbative triples correction Λ-CCSD(T) (ref. 38). The neutron separation energies ($S_n$) of $^{48}$Ca and $^{49}$Ca are computed with the particle-removed/attached equation-of-motion coupled-cluster method truncated at the one-particle-two-hole/two-particle-one-hole excitation level[39]. The three-point mass difference, $\Delta = (S_n(^{48}\text{Ca}) - S_n(^{49}\text{Ca}))/2$, is computed as the difference between two separation energies. The similarity transformed Hamiltonian is non-Hermitian and we compute its right ($|R_0\rangle$) and left ($\langle L_0|$) ground states. Expectation values of one- and two-body operators (*O*) are then obtained from $\langle O \rangle = \langle L_0 | e^{-T} O e^{T} | R_0 \rangle$. In this work we truncate $|R_0\rangle$ and $\langle L_0|$ at the CCSD level. One- and two-body density matrices are computed in a similar fashion. For the computation of the electric dipole polarizability ($\alpha_D$) we used the Lorentz integral transform combined with the coupled-cluster method to properly take the continuum into account[40].

**Computation of intrinsic (weak-) charge densities and radii.**

For the computation of the point-neutron ($R_\mathrm{n}$) and point-proton radii ($R_\mathrm{p}$) we start from the intrinsic operators $R_\mathrm{p}^2 = \frac{1}{Z}\sum_{i=1}^{A}(r_i - R_\mathrm{cm})^2 \left(\frac{1+\tau_i^3}{2}\right)$ and $R_\mathrm{n}^2 = \frac{1}{N}\sum_{i=1}^{A}(r_i - R_\mathrm{cm})^2 \left(\frac{1-\tau_i^3}{2}\right)$. Here $A$ is the number of nucleons, $Z$ the number of protons, $N$ the number of neutrons, $R_\mathrm{cm}$ is the center-of-mass coordinate, and $\tau_i^3$ is the third component of the isospin of the $i^\mathrm{th}$ nucleon. Since $R_\mathrm{p,n}^2$ is a two-body operator, we compute its expectation value by employing the two-body density matrix in the CCSD approximation. For the intrinsic point-proton and point-neutron densities we first compute the corresponding one-body densities in the laboratory system at the CCSD level. The coupled-cluster wave function factorizes approximately into an intrinsic part times a Gaussian center-of-mass wave function[41]. A de-convolution with respect to the Gaussian center-of-mass wave function[42] yields the intrinsic one-body density. The intrinsic point-proton and point-neutron form factors are obtained from Fourier transforms of the one-body densities; folding these with the nucleon form factors given in ref. 20 yields the intrinsic (weak-) charge form factors. The Fourier transform of the (weak-) charge form factor yields the corresponding intrinsic (weak-) charge density.

In our *ab initio* calculations we compute the point-proton radius $R_\mathrm{p}$, which is related to the charge radius $R_\mathrm{ch}$ by $R_\mathrm{ch}^2 = R_\mathrm{p}^2 + \langle r_\mathrm{p}^2 \rangle + \frac{N}{Z}\langle r_\mathrm{n}^2 \rangle + \frac{3}{4M^2} + \langle r^2 \rangle_\mathrm{so}$. Here $\langle r_\mathrm{p}^2 \rangle = 0.769$ fm$^2$ is the mean squared charge radius of a single proton, $\langle r_\mathrm{n}^2 \rangle = -0.116$ fm$^2$ is that of a single neutron, $\frac{3}{4M^2} = 0.033$ fm$^2$ is the relativistic Darwin-Foldy correction, and $\langle r^2 \rangle_\mathrm{so}$ is the spin-orbit correction. For $^{48}$Ca we obtain $\langle r^2 \rangle_\mathrm{so} = -0.09$ fm$^2$, which is slightly smaller in magnitude than the relativistic mean-field estimates[43] due to configuration mixing. Similarly the weak charge radius $R_W$ is computed from $R_W^2 = \frac{Z}{Q_W}\left[Q_W^\mathrm{p}(R_\mathrm{p}^2 + \tilde{r}_\mathrm{p}^2)\right] + \frac{N}{Q_W}\left[Q_W^\mathrm{n}(R_\mathrm{n}^2 + \tilde{r}_\mathrm{n}^2)\right] + \langle \tilde{r}^2 \rangle_\mathrm{so}$ (ref. 43). Here $Q_W = NQ_W^\mathrm{n} + ZQ_W^\mathrm{p}$ is the total weak charge of the nucleus; $Q_W^\mathrm{n} = -0.9878$ and $Q_W^\mathrm{p} = 0.0721$ are the neutron and proton weak charges, respectively; $R_\mathrm{p,n}^2$ is the mean square point proton/neutron radius; $\tilde{r}_\mathrm{p}^2 = 2.358$ fm$^2$ and $\tilde{r}_\mathrm{n}^2 = 0.777$ fm$^2$ are the weak mean squared radii of the proton and neutron; and $\langle \tilde{r}^2 \rangle_\mathrm{so}$ is the spin-orbit correction to the weak charge radius. We compute $\langle \tilde{r}^2 \rangle_\mathrm{so}$ using the coupled-cluster method in the CCSD approximation and we obtain $\langle \tilde{r}^2 \rangle_\mathrm{so} \approx 0.07$ fm$^2$ for all chiral interactions considered in this work. This is comparable to the relativistic mean-field (RMF) estimate $\langle \tilde{r}^2 \rangle_\mathrm{so} \approx 0.077$ fm$^2$ of ref. 42. Extended Data Figure 1 shows the correlation between the weak charge radius and the point-proton radius of $^{48}$Ca. Extended Data Table 2 summarizes the computed binding energies, one-neutron separation energies, three-point mass differences, electric charge radii, weak charge radii, symmetry energy of the nuclear equation of state, and the slope of the symmetry energy at saturation density for the chiral interactions considered in this work.

**Estimating uncertainties.**

Theoretical errors stem from uncertainties in the input (i.e. the employed Hamiltonian) and the computational method used to solve the quantum many-body problem (e.g. truncations of the coupled-cluster method to low-rank particle-hole excitations and finite configuration spaces). The systematic uncertainties of the employed Hamiltonians are the most difficult to quantify. In this work we gauge them by using a set of six state-of-the-art interactions and by correlating the computed observables. Method uncertainties are estimated from benchmark calculations. Benchmark results[23] for ⁴He show that coupled-cluster calculations in the CCSD approximation yield an intrinsic radius that is by about 1% too large when compared to numerically exact calculations from configuration interaction. Coupled-cluster theory is size-extensive, and we assume that radii computed for heavier nuclei (for example $^{40,48}$Ca) similarly exhibit an uncertainty of about 1%. Regarding the uncertainty due to the truncation of the model-space, we find that the point-nucleon radii in ⁴⁸Ca increase by 0.02 fm when increasing the model space from $E_{3\text{max}} = 14\hbar\omega$ to $E_{3\text{max}} = 16\hbar\omega$. It is expected that increasing the model-space size beyond the current limit will slightly increase the computed radii. Our CCSD computations overestimate the radii slightly, thus compensating for part of the model space uncertainty. We thereby arrive at a total method uncertainty of about 1% coming from both the CCSD approximation and the model-space truncation. We also verified that the CCSD result for the electric dipole polarizability α_D for ⁴He is within 1% of the numerically exact hyper-spherical harmonics approach. Combining this uncertainty with the model space truncation we arrive at an uncertainty estimate of 2% for α_D in ⁴⁸Ca. These method uncertainties are shown as error bars on the computed data in Figs. 2, 3 and 4. The blue lines of Figs. 2 and 4 are linear least squares fits to the computed data points. The blue bands encompass the error bars on the computed data points and are chosen symmetrically around the blue line.

**Nuclear Density Functional Theory results.**
The DFT results used in this work were obtained in refs. 2,19 using the energy density functionals SkM*, SkP, SLy4, SV-min, UNEDF0, and UNEDF1.

**Computation of nuclear equation of state from chiral interactions and constraints on neutron-star radii.**
The energy per particle of asymmetric nuclear matter is calculated in many-body perturbation theory up to second order as a function of the neutron and proton densities $\rho_n$ and $\rho_p$ for general isospin asymmetries $\beta = \frac{\rho_n - \rho_p}{\rho}$ (ref. 44). Here $\rho = \rho_n + \rho_p$ denotes the total particle density. In order to extract the values for the symmetry energy parameters $S_v = \frac{1}{2}\partial_\beta^2 E(\beta,\rho)/A|_{\beta=0,\rho=\rho_s}$ and $L = 3\rho_s \partial_\rho S_v(\rho)|_{\rho=\rho_s}$ at the calculated saturation density $\rho_s$, we fit the energy per particle for each Hamiltonian globally in form of a power series in the density and isospin asymmetry. These fits reproduce the calculated microscopic results to high precision and allow us to calculate all relevant observables analytically. For the calculation of neutron star matter we first determine the proton fraction in beta equilibrium by minimizing the nuclear energy plus the energy of a free ultra-relativistic electron gas with respect to the isospin asymmetry. For applications to neutron stars we determine the pressure, $P(\beta,\rho) = \rho^2 \partial_\rho E(\beta,\rho)/A$, at this proton fraction and at the total density $\rho = 0.16$ fm⁻³. In ref. 31 it was shown that the radius $R$ of a neutron star of mass $M$ is tightly correlated with the pressure $P(\rho)$ via the empirical relation $R(M) = C(\rho,M)(P(\rho)/\text{MeV fm}^{-3})^{1/4}$, whereas the value of the parameter $C$ has been constrained to $C(\rho = 0.16 \text{ fm}^{-3}, M = 1.4 M_\odot) = 9.52 \pm 0.49$ km (ref. 30) based on a set of equations of state that support a neutron star with two solar masses. Extended Data Figure 3 shows the correlation between the computed pressure of neutron-star matter at saturation density and the point-proton radius of ⁴⁸Ca. From this correlation and the precisely known charge radius of ⁴⁸Ca we can obtain the pressure of neutron-star matter at $\rho = 0.16$ fm$^{-3}$ and in turn the radius $R_{1.4M_\odot}$ for a neutron star of mass $1.4M_\odot$ (see Fig. 4c).

**Status of *ab-initio* computations.**
Figure 1a is based on refs. 23,45-48.

Figure 1b shows the trend of realistic *ab initio* computations, i.e., *ab initio* computations employing nucleon-nucleon and three-nucleon forces that yield binding energies that agree with experimental data within about 5% or better. It is based on refs. 23,49-61. Calculations for $^{48}$Ca were carried out in this work.

33. S. K. Bogner, R. J. Furnstahl, R. J. Perry, Similarity renormalization group for nucleon-nucleon interactions, *Phys. Rev. C* **75**, 061001(R) (2007).

34. D. R. Entem, R. Machleidt, Accurate charge-dependent nucleon-nucleon potential at fourth order of chiral perturbation theory, *Phys. Rev. C* **68**, 041001(R) (2003).

35. J. D. Holt, J. Menéndez, J. Simonis, A. Schwenk, Three-nucleon forces and spectroscopy of neutron-rich calcium isotopes, *Phys. Rev. C* **90**, 024312 (2014).

36. G. Hagen *et al.*, Coupled-cluster theory for three-body Hamiltonians, *Phys. Rev. C* **76**, 034302 (2007).

37. R. Roth *et al.*, Medium-mass nuclei with normal-ordered chiral NN+3N interactions, *Phys. Rev. Lett.* **109**, 052501 (2012).

38. A. G. Taube, R. J. Bartlett, Improving upon CCSD(T): ΛCCSD(T). I. Potential energy surfaces, *J. Chem. Phys*. **128**, 044110 (2008).

39. J. R. Gour, P. Piecuch, M. Hjorth-Jensen, M. Włoch, D. J. Dean, Coupled-cluster calculations for valence systems around $^{16}$O, *Phys. Rev. C* **74**, 024310 (2006).

40. S. Bacca, N. Barnea, G. Hagen, G. Orlandini, T. Papenbrock, First principles description of the giant dipole resonance in $^{16}$O, *Phys. Rev. Lett.* **111**, 122502 (2013).

41. G. Hagen, T. Papenbrock, D. J. Dean, Solution of the center-of-mass problem in nuclear structure calculations, *Phys. Rev. Lett.* **103**, 062503 (2009).

42. R. Kanungo et al, Exploring the anomaly in the interaction cross section and matter radius of $^{23}$O, *Phys. Rev. C* **84**, 061304(R) (2011).

43. C. J. Horowitz, J. Piekarewicz, Impact of spin-orbit currents on the electroweak skin of neutron-rich nuclei, *Phys. Rev. C* **86**, 045503 (2012).

44. C. Drischler, V. Somà, A. Schwenk, Microscopic calculations and energy expansions for neutron-rich matter, *Phys. Rev. C* **89**, 025806 (2014).

45. U. van Kolck, Few-nucleon forces from chiral Lagrangians, *Phys. Rev. C* **49**, 2932 (1994).

46. E. Epelbaum *et al.,* Three-nucleon forces from chiral effective field theory, *Phys. Rev. C* **66**, 064001 (2002).

47. D. R. Entem, R. Machleidt, Accurate nucleon–nucleon potential based upon chiral perturbation theory, *Phys. Lett. B* **524**, 93 (2002).

48. V. Bernard, E. Epelbaum, H. Krebs, Ulf-G. Meißner, Subleading contributions to the chiral three-nucleon force. II. Short-range terms and relativistic corrections, *Phys. Rev. C* **84**, 054001 (2011).

49. C. R. Chen, G. L. Payne, J. L. Friar, B. F. Gibson, Convergence of Faddeev partial-wave series for triton ground state, *Phys. Rev. C* **31**, 2266 (1985).

50. J. Carlson, Green's function Monte Carlo study of light nuclei, *Phys. Rev. C* **36**, 2026 (1987).

51. B. S. Pudliner, V. R. Pandharipande, J. Carlson, R. B. Wiringa, Quantum Monte Carlo calculations of A≤6 Nuclei, *Phys. Rev. Lett.* **74**, 4396 (1995).

52. R. B. Wiringa, Steven C. Pieper, J. Carlson, V. R. Pandharipande, Quantum Monte Carlo calculations of A=8 nuclei, *Phys. Rev. C* **62**, 014001, (2000).

53. B. Mihaila, J.H. Heisenberg, Microscopic calculation of the inclusive electron scattering structure function in $^{16}$O, *Phys. Rev. Lett.* **84**, 1403 (2000).

54. S. C. Pieper, K. Varga, R. B. Wiringa, Quantum Monte Carlo calculations of A=9,10 nuclei, *Phys. Rev. C* **66**, 044310 (2002).

55. P. Navrátil, V. G. Gueorguiev, J. P. Vary, W. E. Ormand, A. Nogga, Structure of A=10–13 nuclei with two- plus three-nucleon interactions from chiral effective field theory, *Phys. Rev. Lett.* **99**, 042501 (2007).

56. P. Maris *et al.*, Origin of the anomalous long lifetime of $^{14}$C, *Phys. Rev. Lett.* **106**, 202502 (2011).

57. H. Hergert, S. Binder, A. Calci, J. Langhammer, R. Roth, Ab Initio Calculations of Even Oxygen Isotopes with Chiral Two-Plus-Three-Nucleon Interactions, *Phys. Rev. Lett.* **110**, 242501 (2013).

58. A. Cipollone, C. Barbieri, P. Navrátil, Isotopic chains around oxygen from evolved chiral two- and three-nucleon interactions, *Phys. Rev. Lett.* **111**, 062501 (2013).

59. S. K. Bogner *et al.*, Nonperturbative shell-model interactions from the in-medium similarity renormalization group, *Phys. Rev. Lett.* **113**, 142501 (2014).

60. G. R. Jansen, J. Engel, G. Hagen, P. Navrátil, A. Signoracci, Ab initio coupled-cluster effective interactions for the shell model: application to neutron-rich oxygen and carbon isotopes, *Phys. Rev. Lett.* **113**, 142502 (2014).

61. T. Lähde *et al.*, Lattice effective field theory for medium-mass nuclei, *Phys. Lett. B* **732**, 110 (2014).

62. M. C. M. Rentmeester, R. G. E. Timmermans, J. J. de Swart, Determination of the chiral coupling constants $c_3$ and $c_4$ in new pp and np partial-wave analyses, *Phys. Rev. C* **67**, 044001 (2003).

**Acknowledgments** We acknowledge discussions with C. Horowitz, J. Piekarewicz, P.-G. Reinhard, and A. Steiner. This material is based upon work supported by the U.S. Department of Energy, Office of Science, Office of Nuclear Physics under Award Numbers DEFG02-96ER40963 (University of Tennessee), DOE-DE-SC0013365 (Michigan State University), DE-SC0008499 and DE-SC0008511 (NUCLEI SciDAC collaboration), the Field Work Proposal ERKBP57 at Oak Ridge National Laboratory and the National Science Foundation with award number 1404159. It was also supported by the Swedish Foundation for International Cooperation in Research and Higher Education (STINT, IG2012-5158), by the European Research Council (ERC-StG-240603), by NSERC Grant No. 2015-00031, by the ERC Grant No. 307986 STRONGINT, and the Research Council of Norway under contract ISPFysikk/216699. TRIUMF receives funding via a contribution through the National Research Council Canada. Computer time was provided by the INCITE program. This research used resources of the Oak Ridge Leadership Computing Facility located at Oak Ridge National Laboratory, which is supported by the Office of Science of the Department of Energy under Contract No. DEAC05-00OR22725; and computing resources at the Jülich Supercomputing Center.

**Author contributions** G.H. initiated and led the project. G.H., A.E., G.R.J, T.P., K.A.W., S.B., N.B., B.C., C.D., K.H., M.H.J., M.M., G.O., A.S., and J.S. developed computational tools utilized in this study. G.H., G.R.J, K.A.W., C.D., K.H., and M.M. performed calculations. G.H., A.E., C.F., G.R.J, W.N., T.P., K.A.W., S.B., N.B., C.D., K.H., M.H.J., M.M., G.O., and A.S. discussed and interpreted the results. G.H, A.E, C.F, G.R.J., W.N, T.P., K.A.W, K.H., A.S. wrote the paper with input from all co-authors.

**Competing financial interests** The authors declare no competing financial interests.

**Author information** Correspondence and requests for materials should be addressed to hageng@ornl.gov

**Extended Data**

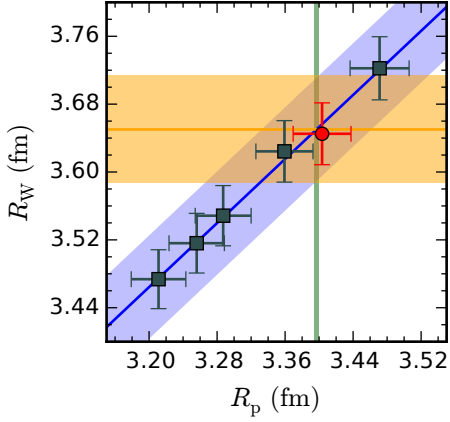

**Extended Data Figure 1 | Correlation between weak and point-proton radius in $^{48}$Ca.** Correlation between the weak charge radius $R_W$ and the point-proton radius $R_p$ for $^{48}$Ca from the chiral interactions used in this work. The red dot denotes the NNLO$_{sat}$ result while the results of chiral interactions listed in Extended Data Table 1 are marked by dark squares. The theoretical error bars are indicated. The blue line represents a linear fit to the data, with theoretical uncertainties shown by a blue band. The vertical green line marks the experimental value of $R_p$ that places a constraint on the weak charge radius $R_W$ (horizontal orange band).

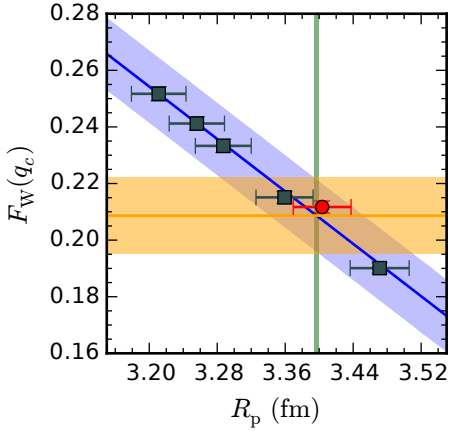

**Extended Data Figure 2 | Correlation between weak form-factor and point-proton radius in $^{48}$Ca.** Similar as in Extended Data Fig. 1 but for the weak form factor $F_W(q_c)$ at the CREX momentum $q_c = 0.778$ fm$^{-1}$ and the point-proton radius $R_p$ in $^{48}$Ca.

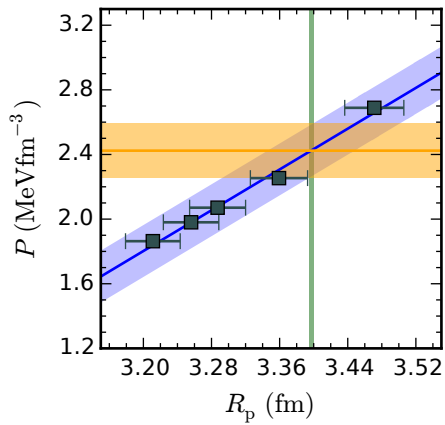

**Extended Data Figure 3 | Correlation between pressure of neutron star matter and the point-proton radius of $^{48}$Ca.** Similar as in Extended Data Fig. 1 but for the pressure $P$ of neutron-star matter at the empirical saturation density $\rho_s=0.16$ fm$^{-3}$ and the point-proton radius $R_p$ in $^{48}$Ca. NNLO$_{sat}$ results are excluded because this interaction was constrained only by low-energy data and is not expected to work at neutron densities around 0.16 fm$^{-3}$.

**Extended Data Table 1 | Parameters of chiral interactions.** List of chiral interactions (excluding NNLO$_{sat}$) used in this work. Here $\Lambda_{SRG}$ (in fm$^{-1}$) is the momentum scale of the similarity renormalization group transformation (SRG) used to soften the nucleon-nucleon interaction; $\Lambda_{3NF}$ (in fm$^{-1}$) is the cutoff in the non-local regulator of the three-nucleon force; $c_D$, and $c_E$ (dimensionless) are the low-energy constants of the short-range terms of the three-nucleon force. EM indicates that low-energy constants ($c_i$) for the long-range terms (two-pion exchange) are taken from ref. 34, and PWA refers to $c_i$ values from ref. 62.

| Label | $\Lambda_{SRG}$ | $\Lambda_{3NF}$ | $c_D$ | $c_E$ |
|---|---|---|---|---|
| 1.8/2.0 (EM) | 1.8 | 2.0 | +1.264 | -0.120 |
| 2.0/2.0 (EM) | 2.0 | 2.0 | +1.271 | -0.131 |
| 2.2/2.0 (EM) | 2.2 | 2.0 | +1.214 | -0.137 |
| 2.8/2.0 (EM) | 2.8 | 2.0 | +1.278 | -0.078 |
| 2.0/2.0 (PWA) | 2.0 | 2.0 | -3.007 | -0.686 |

**Extended Data Table 2 | Key observables from chiral interactions.** Predictions for $^{48}$Ca (based on the interactions used in this work): binding energy $BE$, neutron separation energy $S_n$, three-point-mass difference $\Delta$, electric-charge radius $R_{ch}$, and the weak-charge radius $R_W$. The last two columns show the symmetry energy of the nuclear equation of state and its slope $L$ at saturation density. Energies are in MeV and radii in fm. Theoretical uncertainty estimates are about 1% for radii and energies.

| Interaction | $BE$ | $S_n$ | $\Delta$ | $R_{ch}$ | $R_W$ | $S_v$ | $L$ |
|---|---|---|---|---|---|---|---|
| NNLO$_{sat}$ | 404 | 9.5 | 2.69 | 3.48 | 3.65 | 26.9 | 40.8 |
| 1.8/2.0 (EM) | 420 | 10.1 | 2.69 | 3.30 | 3.47 | 33.3 | 48.6 |
| 2.0/2.0 (EM) | 396 | 9.3 | 2.66 | 3.34 | 3.52 | 31.4 | 46.7 |
| 2.2/2.0 (EM) | 379 | 8.8 | 2.61 | 3.37 | 3.55 | 30.2 | 45.5 |
| 2.8/2.0 (EM) | 351 | 8.0 | 2.41 | 3.44 | 3.62 | 28.5 | 43.8 |
| 2.0/2.0 (PWA) | 346 | 7.8 | 2.82 | 3.55 | 3.72 | 27.4 | 44.0 |
| Experiment | 415.99 | 9.995 | 2.399 | 3.477 | | | |